\documentclass[aps,twocolumn,showpacs,prl]{revtex4}
\usepackage{graphicx}
\newlength{\figwidth}
\begin{document}

\title{Mott transition in Kagom\'e lattice Hubbard model}

\author{Takuma Ohashi}
\author{Norio Kawakami}
\affiliation{%
Department of Applied Physics, Osaka University, 
Suita, Osaka 565-0871, Japan}%

\author{Hirokazu Tsunetsugu}
\affiliation{%
Yukawa Institute for Theoretical Physics, 
Kyoto University, Kyoto 606-8502, Japan}%

\date{\today}

\begin{abstract}
We investigate the Mott transition in the Kagom\'e lattice Hubbard model
using a cluster extension of dynamical mean field theory. 
The calculation of the double occupancy, the density of states, 
the static and dynamical spin correlation functions demonstrates 
that the system undergoes the first-order Mott transition 
at the Hubbard interaction $U/W \sim 1.4$ ($W$:bandwidth). 
In the metallic phase close to the Mott transition, 
we find the strong renormalization of three distinct bands, 
giving rise to the formation of heavy quasiparticles with strong frustration.
It is elucidated that the quasiparticle states 
exhibit anomalous behavior in the temperature-dependent spin correlation
functions. 
\end{abstract}

\maketitle

Geometrically frustrated electron systems have provided hot 
topics in the field of strongly correlated electron systems. 
The observation of heavy fermion behavior in $\mathrm{LiV_2O_4}$
\cite{kondo97}, which 
has the pyrochlore lattice structure with a corner-sharing
network of tetrahedra, has activated 
theoretical studies of electron correlations with
geometrical frustration.
The discovery of superconductivity in the triangular lattice compound
$\mathrm{Na_xCoO_2 \cdot yH_2O}$ \cite{takada03}
and the $\beta$-pyrochlore osmate $\mathrm{KOs_2O_6}$ \cite{yonezawa04}
has further stimulated intensive studies of frustrated electron systems. 
Geometrical frustration has uncovered new aspects
of the Mott metal-insulator transition, 
which is now one of the central issues in the physics of 
strongly correlated electron systems. Among others, 
a novel quantum liquid ground state suggested 
for the Mott insulating phase of the triangular lattice \cite{kashima01}
may be relevant for frustrated organic materials such as 
$\kappa$-$\mathrm{(ET)_2Cu_2(CN)_3}$ \cite{shimizu03}.

\begin{figure}[bt]
\begin{center}
\includegraphics[clip,width=0.45\textwidth]{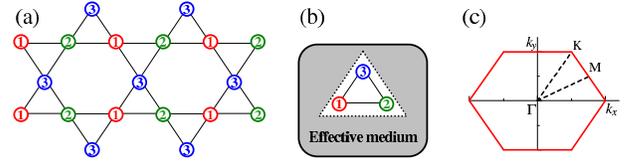}
\end{center}
\caption{
(a) Sketch of the Kagom\'e lattice and 
(b) the effective cluster model using three sites cluster CDMFT. 
(c) First Brillouin zone of the Kagom\'e lattice. 
}
\label{fig:kagome}
\end{figure}

The Kagom\'e lattice (Fig. \ref{fig:kagome}) is another prototype of 
 frustrated systems, 
which may be regarded as a two-dimensional analog of 
the pyrochlore lattice. It is suggested 
that a correlated electron system on the Kagom\'e lattice 
can be an effective model of $\mathrm{Na_xCoO_2 \cdot yH_2O}$ 
by properly considering anisotropic hopping matrix elements 
in the cobalt 3$d$ orbitals \cite{koshibae03}. 
The issue of electron correlations for the Kagom\'e lattice
was addressed recently by using the FLEX approximation \cite{imai03} 
and QMC method \cite{bulut05}. 
These studies focused on electron correlations in the metallic regime, 
and the nature of the Mott transition has not been clarified yet. 
Therefore, it is desirable to investigate the
 Kagom\'e lattice electron system with particular emphasis on
 the Mott transition under the influence of strong frustration.

In this paper, we study the Mott transition of correlated electrons 
on the Kagom\'e lattice by means of the cellular dynamical 
mean field theory (CDFMT) \cite{kotliar01}. It is shown that the 
metallic phase persists up to fairly large Coulomb interactions 
due to the frustrated lattice structure.
This gives rise to the strong renormalization of three distinct bands,
resulting in the multi-band quasiparticles with strong 
frustration near the Mott transition.  In particular, we 
find that the quasiparticles exhibit 
anomalous behavior in spin correlation functions, which 
characterizes strong frustration in the metallic phase. 

We consider the standard Hubbard model with nearest-neighbor 
hopping on the Kagom\'e lattice,
\begin{eqnarray}
H= -t \sum_{\left \langle i,j \right \rangle ,\sigma}
c_{i\sigma }^\dag c_{j\sigma}
+ U \sum_{i} n_{i\uparrow} n_{i\downarrow} \ (t>0), 
\label{eqn:hm}
\end{eqnarray}
with $n_{i\sigma}=c_{i\sigma}^\dag c_{i\sigma}$, 
where $c_{i\sigma }^\dag$ ($c_{j\sigma}$) creates (annihilates) 
an electron with spin $\sigma$ at site $i$.
We use the band width $W=6t$ as the energy unit. 
To study the Mott transition in the Kagom\'e lattice system,
we need an efficient theoretical tool to treat both of 
strong correlations and geometrical frustration. 
The dynamical mean field theory (DMFT) 
\cite{georges96} has given substantial theoretical 
progress in the field of the Mott transition 
but it does not incorporate spatially extended correlations.
In order to treat both strong correlations and 
frustration, we use CDMFT, a cluster extension of DMFT, 
which has been successfully applied to frustrated systems such as 
the Hubbard model on the triangular lattice \cite{parcollet04}. 

In CDMFT, the original lattice is regarded 
as a superlattice consisting of clusters, which
is then mapped onto an effective cluster model via a standard DMFT procedure. 
Each unit cell of the Kagom\'e lattice has three sites labeled by 
$1$, $2$, and $3$, as shown in Fig. \ref{fig:kagome}(a). 
We thus end up with a three-site cluster model 
coupled to the self-consistently determined medium 
illustrated in Fig. \ref{fig:kagome}(b). 
Given the Green's function for the effective medium, 
$\hat{\mathcal{G}}_{\sigma}$,
we can compute the cluster Green's function $\hat{G}_{\sigma}$
and the cluster self-energy $\hat{\Sigma}_{\sigma}$ 
by solving the effective cluster model with QMC method \cite{hirsch86}. 
Here, $\hat{\mathcal{G}}_{\sigma}$, $\hat{G}_{\sigma}$, 
and $\hat{\Sigma}_{\sigma}$ are described by 
$3 \times 3$ matrices. 
The effective medium $\hat{\mathcal{G}}_{\sigma}$ 
is then computed via the Dyson equation, 
\begin{eqnarray}
\hat{\mathcal{G}}_{\sigma}^{-1} \left( \omega \right) = 
\left [ \sum_\mathbf{K}
  \frac{1}{\omega + \mu - \hat{t} \left( \mathbf{K} \right)
  - \hat{\Sigma}_\sigma \left( \omega \right)}
\right ] ^{-1}
+ \hat{\Sigma}_{\sigma} \left( \omega \right) ,
\nonumber \\
\label{eqn:cavity}
\end{eqnarray}
where $\mu$ is the chemical potential. 
Here the summation of $\mathbf{K}$ is taken over the reduced Brillouin zone 
of the superlattice (see Fig. \ref{fig:kagome}(c)) 
and $\hat{t} \left( \mathbf{K} \right)$ is the 
Fourier-transformed hopping matrix for the superlattice. 
After twenty times iteration of this procedure, 
numerical convergence is reached. In each iteration, 
we typically use $10^6$ QMC sweeps and 
Trotter time slices $L = 2W/T$ 
to reach sufficient computational accuracy. 
Furthermore, 
we exploit an interpolation scheme based on a high-frequency expansion
of the discrete imaginary-time Green's function obtained by QMC
\cite{oudovenko02} in order to reduce numerical errors resulting from 
the Fourier transformation from imaginary time to Matsubara frequency. 

\begin{figure}[bt]
\begin{center}
\includegraphics[clip,width=\figwidth]{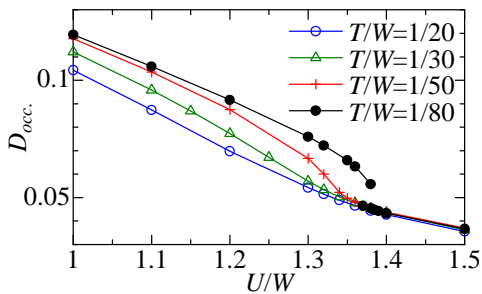}
\end{center}
\caption{
Double occupancy as a function of interaction strength $U/W$ 
for several temperatures $T/W$. At $T/W=1/80$, we can see 
the discontinuity with hysteresis, indicating the first-order
Mott transition.
}
\label{fig:double}
\end{figure}

Let us now investigate the Mott transition 
of the Kagom\'e lattice Hubbard model at half filling. 
In Fig. \ref{fig:double}, we show the results for the double occupancy 
$D_{occ.} = \left \langle n_{i\uparrow} n_{i\downarrow} \right \rangle$ 
at various temperatures. At high temperatures, $D_{occ.}$ 
smoothly decreases as $U$ increases, indicating the development 
of local spin moments. As the temperature
is lowered, there appears singular behavior around 
 characteristic values of $U$.
When $1/50 \le T/W \le 1/20$, $D_{occ. }$ shows crossover behavior at 
$U/W \sim 1.35$. 
At lower temperature $T/W = 1/80$, the crossover is changed to 
the discontinuity accompanied by hysteresis, which signals a 
first-order phase transition at $U_c/W \sim 1.37$. 
This is the first demonstration of the Mott transition in 
the Kagom\'e lattice Hubbard model. 
Note that the critical interaction strength $U_c$ is much larger than 
the crossover strength of $U$ found for the unfrustrated square 
lattice \cite{moukouri01}. 
As is the case for the triangular-lattice Hubbard 
model \cite{parcollet04},   
the double occupancy $D_{occ.}$ increases in the metallic 
phase ($U<U_c$) as $T$ decreases,
while it is almost independent of $T$ in the insulating 
phase ($U>U_c$). The increase of $D_{occ.}$ at low temperatures
means the suppression of the local moments due to the
itinerancy of electrons, in other words, the formation
of quasiparticles. Note that 
in the metallic phase close to the transition point,
the increase of $D_{occ.}$ occurs at very low temperatures. 
This implies that the coherence temperature $T_C$ that characterizes
the formation of quasiparticles is very low.
This naturally causes strong frustration and,
as shown below, brings about unusual metallic properties near 
the Mott transition.

\begin{figure}[bt]
\begin{center}
\includegraphics[clip,width=\figwidth]{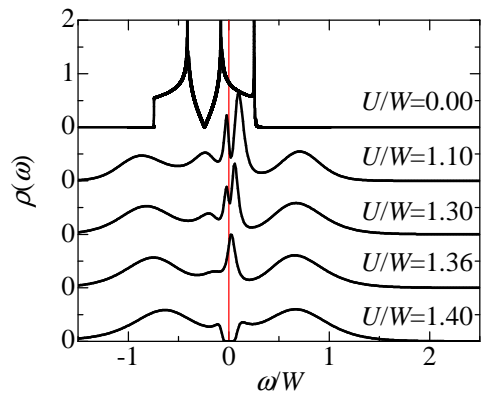}
\end{center}
\caption{
Density of states at $T/W=1/80$ for several strengths of $U/W$. 
}
\label{fig:dos}
\end{figure}

To see how the quasiparticles evolve around the Mott 
transition point clearly, 
we calculate the density of states (DOS) by applying
the maximum entropy method (MEM) \cite{jarrell96} 
to the imaginary-time QMC data. 
In Fig. \ref{fig:dos}, we show DOS at $T/W=1/80$ for 
several values of the interaction $U/W$. 
In the non-interacting case ($U=0$), DOS has 
three distinct bands including a $\delta$-function peak
above the Fermi level.
With increasing $U/W$, the DOS forms
heavy quasiparticle peaks around the Fermi level
and eventually develops a dip at $U/W \sim 1.40$,
signaling the Mott transition.
There are two characteristic properties in the metallic phase 
close to the critical point. First we note that
heavy quasiparticles persist up to the 
transition point ($U/W=1.30$ and $1.36$) and
there is no evidence for the pseudo-gap formation, which 
 is consistent with the $U$- and $T$-dependence of 
the double occupancy in Fig. \ref{fig:double}.
This is related to the suppression of magnetic instabilities 
in our system, in contrast to the square lattice case,
where the quasiparticle states are strongly suppressed and 
a pseudo gap opens. 
Another point to be noticed is how strongly
 the renormalization occurs near the critical point.
One can see three renormalized peaks near the 
Fermi level: not only the electrons near the Fermi surface 
 but also the two bands away from the Fermi surface are renormalized to
participate in the formation of quasiparticles. 
Therefore, the three quasiparticle bands are all relevant
for low-energy excitations near the Mott transition, in contrast to
the weak coupling regime where only the single band 
around the Fermi surface is relevant. 

\begin{figure}[bt]
\begin{center}
\includegraphics[clip,width=\figwidth]{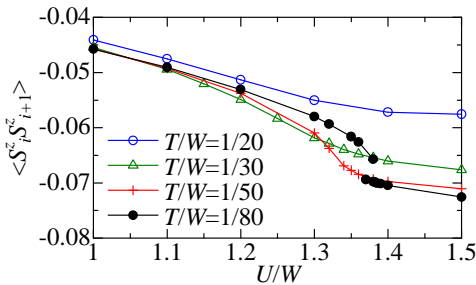}
\end{center}
\caption{
The nearest neighbor spin correlation function 
$\langle S^z_i S^z_{i+1} \rangle$ as a function of $U/W$ 
at several temperatures. 
Note that the low-temperature spin correlation in the insulating phase is
somewhat weaker than that for the isolated triangle, 
$\langle S^z_i S^z_{i+1} \rangle = -1/12$. 
}
\label{fig:correlation}
\end{figure}

The remarkable fact we find is that 
the quasiparticles show anomalous properties in
spin correlations due to strong
frustration
 around the transition point. Shown in Fig. \ref{fig:correlation} is
the nearest-neighbor spin correlation function 
$\langle S^z_i S^z_{i+1} \rangle$ at different temperatures. 
$\langle S^z_i S^z_{i+1} \rangle$ is always negative so that 
the spin correlation is antiferromagnetic (AF), 
which is  a source of strong frustration. 
As $U/W$ increases, the nearest-neighbor AF spin correlation 
is enhanced gradually. 
In the insulating phase 
the AF spin correlation gets stronger with decreasing temperature, 
as is expected. 
We can see that more striking behavior
emerges in the metallic phase close to the critical point: 
the AF spin correlation is once enhanced and then suppressed 
with the decrease of the temperature. 
The anomalous temperature dependence results from
 the competition between the quasiparticle formation 
and the frustrated spin correlations, which 
may be characterized by two energy scales: 
the coherence temperature $T_C$ and 
$T_M$ characterizing the AF spin fluctuations. 
The AF correlation is developed around $T \sim T_M$, 
which stabilizes localized moments and causes frustration
in accordance with the monotonic enhancement 
of spin correlations in the insulating phase in Fig. \ref{fig:correlation}.
On the other hand, 
when the system is in the metallic phase, 
electrons recover coherence in itinerant motion below $T_C$. 
Therefore, the frustration is relaxed by itinerancy of electrons
via the suppression of AF correlations at $T<T_C$. 
Thus, the nonmonotonic temperature-dependence of
$\langle S^z_i S^z_{i+1} \rangle$ clearly demonstrates that 
the heavy quasiparticles are formed 
under the influence of strong frustration. 

\begin{figure}[bt]
\begin{center}
\includegraphics[clip,width=\figwidth]{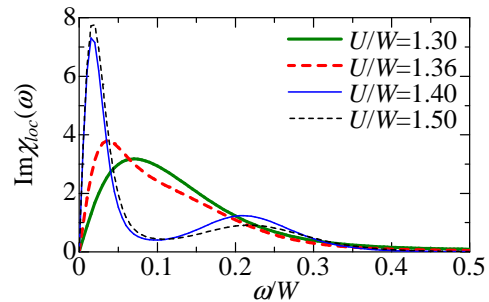}
\end{center}
\caption{
Dynamical susceptibility $\mathrm{Im} \chi _{loc} ( \omega )$ 
at $T/W=1/80$ for several strengths of $U/W$. 
}
\label{fig:spin}
\end{figure}

The anomalous properties also appear 
in dynamical spin correlation functions. 
We calculate the dynamical spin susceptibility 
$\chi _{loc} ( \omega ) = - i \int dt e^{i \omega t}
\langle [ S^z_i (t), S^z_i(0)] \rangle$, where 
$S^z_i=( c_{i\uparrow}^\dag c_{i\uparrow} 
- c_{i\downarrow}^\dag c_{i\downarrow} )/2$. 
In Fig. \ref{fig:spin}, we show $\mathrm{Im} \chi _{loc} ( \omega )$ 
around the Mott transition at $T/W=1/80$. 
A remarkable point is that 
the profile of $\mathrm{Im} \chi _{loc} ( \omega )$ 
dramatically changes around the Mott transition. 
In the insulating phase ($U/W=1.4$, $1.5$), 
$\mathrm{Im} \chi _{loc} ( \omega )$ has 
two distinct peaks at low energies. 
In the metallic phase ($U/W=1.3$, $1.36$), 
two peaks get renormalized into a single peak 
and its peak value is strongly suppressed. 
This is the first demonstration of drastic change of spin dynamics 
between metallic and insulating phases in frustrated systems. 
In the insulating phase, 
the short-range AF correlations become dominant at low temperatures, 
resulting in the appearance of the double-peak structure. 
The strongly enhanced low-energy peak in $\chi _{loc} ( \omega )$ 
corresponds to excitations among the almost degenerate states 
for which a singlet spin pair is formed inside the unit cell, 
while the higher-energy hump is caused by the excitations 
from these low-energy states to other excited states. 
In the metallic phase, the AF correlations are suppressed 
and then frustration is relaxed via the itinerancy of electrons, 
which leads to the renormalized single peak structure 
in $\chi _{loc} ( \omega )$. 
Therefore, the dramatic change in $\chi _{loc} ( \omega )$ 
features the competition between itinerancy and frustration of
correlated electrons around the Mott transition point. 

\begin{figure}[bt]
\begin{center}
\includegraphics[clip,width=\figwidth]{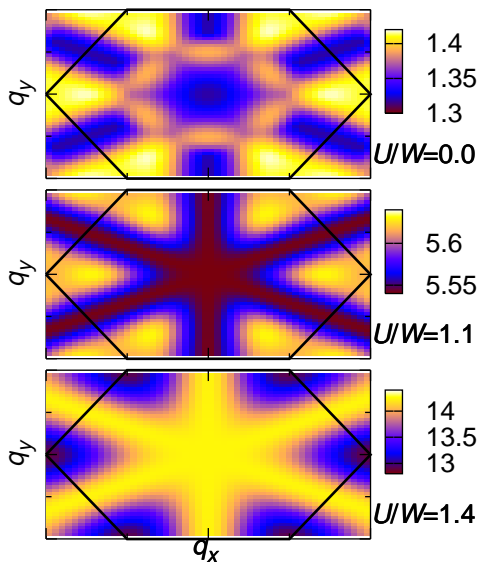}
\end{center}
\caption{
The maximum mode of the susceptibility $\chi_{max}(\mathbf{q})$ 
for different strength of $U/W$ at $T/W=1/30$. 
Hexagons in figures denotes the first Brillouin zone as shown 
Fig. \ref{fig:kagome} (c). 
}
\label{fig:susceptibility}
\end{figure}

Finally, we discuss the magnetic instability by 
examining the wavevector-dependence of the static susceptibility,
\begin{eqnarray}
&&\chi_{\gamma \delta}(\mathbf{q}) = \int_0^{1/T} d \tau 
\sum_{\mathbf{k},\mathbf{k}'} 
\nonumber \\
&&\times
\left \langle 
c_{\mathbf{k}\gamma\uparrow}^\dag               \left ( \tau \right ) 
c_{\mathbf{k}+\mathbf{q}\gamma\downarrow}       \left ( \tau \right )
c_{\mathbf{k}'+\mathbf{q}\delta\downarrow}^\dag \left ( 0 \right )
c_{\mathbf{k}'\delta\uparrow}                   \left ( 0 \right )
\right \rangle ,
\end{eqnarray}
where $\gamma,\delta =1,2,3$ denote the superlattice indices. 
We employ the standard procedure in DMFT to calculate 
$\chi_{\gamma \delta}(\mathbf{q})$ \cite{georges96}, 
which includes nearest-neighbor correlations as well as on-site correlations. 
It is convenient to introduce $\chi_m(\mathbf{q})$ for three normal 
modes ($m=1,2,3$) by diagonalizing the $3 \times 3$ matrix 
$\chi_{\gamma \delta}(\mathbf{q})$. 
We find that the $\mathbf{q}$-dependence of 
the mode with the maximum eigenvalue 
(referred to as $\chi_{max}(\mathbf{q})$) is much weaker than 
that for the other two modes, 
while the second largest mode has the strong 
$\mathbf{q}$-dependence with a maximum at $\mathbf{q}=(0,0)$. 
These results are consistent with the previous FLEX calculation 
\cite{imai03} and the QMC study \cite{bulut05}. 
We find, however, notable new results in the strong coupling regime. 
In Fig. \ref{fig:susceptibility}, we show  $\chi_{max}(\mathbf{q})$ 
for several strengths of $U/W$ at $T/W=1/30$. 
In the noninteracting case $U/W=0$, 
the susceptibility takes a maximum at six points in the Brillouin zone. 
As $U/W$ increases, the susceptibility is enhanced 
not only at these six points but also on the lines 
through $\mathrm{\Gamma}$ and $\mathrm{M}$ points, 
so that $\chi_{max}(\mathbf{q})$ becomes much flatter at $U/W=1.1$ 
than in the noninteracting case. 
Once the system enters the insulating phase, 
the $\mathbf{q}$-dependence of $\chi_{max}(\mathbf{q})$ dramatically changes 
its character due to the enhancement of short range AF correlations. 
At $U/W=1.4$, the susceptibility is further enhanced along 
the three lines in $\mathbf{q}$ space 
and becomes dominant instead of the six points that give 
the leading magnetic mode in the weak coupling regime. 
Furthermore, by investigating the eigenvectors of $\chi_{max}(\mathbf{q})$, 
we find that two spins in the unit cell are antiferromagnetically coupled 
but the other spin is free. 
Therefore, these enhanced spin fluctuations 
favor a spatial spin configuration in which 
one-dimensional AF-correlated spin chains 
are independently formed in three distinct directions. 
This is consistent with the intra chain spin correlations 
in the $\mathbf{q}=0$ structure predicted for the Kagom\'e 
Heisenberg systems \cite{harris92}. 
However, the essential difference is that 
there is almost no correlation between different chains. 
This is the first observation of one-dimensional spin correlations 
in the itinerant electron systems with geometrical frustration. 
Although we have not obtained real instability to a novel 
one-dimensional ordering in the present calculation, 
such enhanced spin fluctuations 
certainly affect low-energy dynamics in the insulating phase. 

In summary we have investigated the Mott transition 
in the Kagom\'e lattice Hubbard model 
using CDMFT combined with QMC. We have found that the 
metallic phase is stabilized up to fairly large $U$,
resulting in the three-band heavy quasiparticles with strong frustration. 
This causes several anomalous properties of spin correlation 
functions in the metallic phase close to the critical point. 
We have also discussed the possibility of magnetic instability 
toward a novel one-dimensional ordering in the insulating phase. 

The authors thank S. Suga, Y. Motome, A. Koga, and Y. Imai 
for valuable discussions. 
A part of numerical computations was done at the Supercomputer Center 
at the Institute for Solid State Physics, University of Tokyo. 
This work was partly supported by a Grant-in-Aid from the Ministry 
of Education, Science, Sports and Culture of Japan.


\begin{thebibliography}{99}
\bibitem{kondo97}
S. Kondo, {\it et al.}, 
Phys. Rev. Lett. {\bf 78}, 3729 (1997). 

\bibitem{takada03}
K. Takada, {\it et al.}, 
Nature (London) {\bf 422}, 53 (2003). 

\bibitem{yonezawa04}
S. Yonezawa, Y. Muraoka, Y. Matsushita, and Z. Hiroi, 
J. Phys.: Cond. Mat. {\bf 16}, L9 (2004). 

\bibitem{kashima01}
T. Kashima, and M. Imada, 
J. Phys. Soc. Jpn. {\bf 70}, 3052 (2001). 

\bibitem{shimizu03}
Y. Shimizu, 
K. Miyagawa, K. Kanoda, M. Maesato, and G. Saito, 
Phys. Rev. Lett. {\bf 91}, 107001 (2003). 

\bibitem{koshibae03}
W. Koshibae and S. Maekawa, 
Phys. Rev. Lett. {\bf 91}, 257003 (2003). 

\bibitem{imai03}
Y. Imai, N. Kawakami, and H. Tsunetsugu
Phys. Rev. B {\bf 68}, 195103 (2003). 

\bibitem{bulut05}
N. Bulut, W. Koshibae, and S. Maekawa
Phys. Rev. Lett. {\bf 95}, 037001 (2005). 

\bibitem{kotliar01}
G. Kotliar, S. Y. Savrasov, G. P\'alsson, and G. Biroli
Phys. Rev. Lett. {\bf 87}, 186401 (2001). 

\bibitem{georges96}
 A. Georges, G. Kotliar, W. Krauth and M. J. Rozenberg, 
 Rev. Mod. Phys. {\bf 68}, 13 (1996). 

\bibitem{parcollet04}
O. Parcollet, G. Biroli, and G. Kotliar, 
Phys. Rev. Lett. {\bf 92}, 226402, (2004). 

\bibitem{hirsch86} 
J. E. Hirsch and R. M. Fye, 
Phys. Rev. Lett. {\bf 56}, 2521 (1986). 

\bibitem{oudovenko02}
V. S. Oudovenko and G. Kotliar
Phys. Rev. B {\bf 65}, 075102 (2002). 

\bibitem{moukouri01}
S. Moukouri, and M. Jarrell
Phys. Rev. Lett. {\bf 87}, 167010, (2001). 

\bibitem{jarrell96}
M. Jarrell and J. E. Gubernatis, 
Phys. Rep. {\bf 269}, 133 (1996). 

\bibitem{harris92}
A. B. Harris, C. Kallin, and A. J. Berlinsky, 
Phys. Rev. B 45, 2899 (1992). 

\end{thebibliography}
\end{document}